\documentclass{aastex}          
\usepackage{spr-astr-addons}    

\begin{document}
%
\title{New Results on Standard Solar Models}

\shorttitle{New Results on Solar Models}
\shortauthors{Aldo Serenelli}

\author{Aldo M. Serenelli} 
\affil{Max  Planck Institute  for  Astrophysics, Karl  Schwarzschild Str.   1,
  Garching, D-85471, Germany} 


\begin{abstract}
We describe the current status of solar modelling and focus on the problems
originated with the introduction of solar abundance determinations with low
CNO abundance values. We use models computed with solar abundance compilations
obtained during the last decade, including the newest published abundances by
Asplund and collaborators. Results presented here make focus both on
helioseismic properties and the models as well as in the neutrino fluxes
predictions. We also discuss changes in radiative opacities to
restore agreement between helioseismology, solar models, and solar abundances
and show the effect of such modifications on solar neutrino fluxes. 
\end{abstract}


\section{Introduction \label{sec:intro}}

Solar models are a corner stone of stellar astrophysics. The determination of
the solar interior structure through helioseismology and, only a few years
later, the discovery that neutrinos change flavor, gave spectacular
confirmations of our ability to model the Sun and, by extension, of other
stars. However, a series of works starting with a redetermination of the
photospheric oxygen solar abundance \citep{allende} and finishing with a
complete revision of solar abundances \citep{ags05}, led to a strong
reduction in the overall metallicity of the Sun driven by much lower CNO and
Ne abundances than previously determined. Soon after, solar models that
adopted the new 
composition were shown to have an interior structure at odds with
helioseismology determinations. Since then, the
so-called {\em solar abundance problem} has been in the spotlight of solar
(and stellar) astrophysics. As nicely put by \citet{dp06}, it represents the
incompatibility between the best solar atmosphere and interior models
available. 

The effects of the low metallicity in the solar interior has been widely
discussed in the literature. Among many other references, the reader can refer
to \citet{basu04,turck04,monta04,bs05,dp06} and \citet{montecarlo}.  Some
attempts to constrain the solar metallicity independently of photospheric
measurements can be found in \citet{antia06,lin07} and \citet{bisonii}.
The connection between solar neutrinos and composition has also been discussed
in different works, e.g. \citet{turck04,bp04,bs05,montecarlo,bps08,wick}.
Possible modifications in the physical inputs of solar models have also been
discussed in connection to the {\em solar abundance problem}. The reader can
refer to \citet{monta04,guzik05,dp06,castro07} just to mention some relevant
works. 

In this article, we present a short review of the field and present new solar
models that incorporate the most recent solar abundance determination by
\citet{agss09}. In \S~\ref{sec:models} we describe the main characteristics of
the models used to obtain the core results presented here, including the
different options for solar compositions we used. Results are presented in 
\S~\ref{sec:results} where helioseismology properties of the models and solar
neutrino fluxes are discussed in the context of current observational and
experimental data. In \S~\ref{sec:opac} we go to some length in discussing
radiative opacities as a possible solution to the abundance problem, including
the effects of opacities in solar neutrino fluxes. We summarize in
\S~\ref{sec:conclu}.

\section{Solar Models\label{sec:models}}

Over the  years, standard solar models  have played a fundamental  role in the
development of stellar astrophysics as well as in the 

Most of the results presented here refer to solar models computed with the 
GARSTEC stellar  evolution code \citep{garstec} with modifications. Some
of the differences are: 
the nuclear energy generation routine is {\em exportenergy.f}\footnote{Publicly 
available at http://www.sns.ias.edu/$\sim$jnb};  radiative opacities are those
from the Opacity Project \citep{op}  complemented at low temperatured by those
from  \citet{lowt}. Unless  stated otherwise,  the  equation of  state (EOS)
is  the revised                               version
of OPAL\footnote{http://adg.llnl.gov/Research/OPAL/EOS\_2005/}. 

As  it is  usual practice  in  solar models,  a 1~M$_\odot$  stellar model  is
evolved from the pre-main sequence (or zero age main sequence) up to the solar
system age (which we take  to be $\tau_\odot$=4.57~\hbox{Gyr}; see appendix in
\citealt{bp95}).  The solar model  is forced  to match  the present  day solar
radius R$_\odot$ and luminosity L$_\odot$ and here we adopt the values
$6.9598 \times 10^{10}$~\hbox{cm}  $3.8418 \times 10^{33}$~\hbox{erg s$^{-1}$}
respectively. The third condition for solar models is to match the present-day
metal to hydrogen fraction $\left(Z/X\right)_{\rm ph}$ in the solar photosphere.
The models presented here have been computed for three different basic sets of
solar abundances as follows:
\begin{itemize}
\item[] GS98: abundances from \citet{gs98} where meteoritic abundances 
  for refractories are adopted and $\left(Z/X\right)_{\rm ph}=0.0229$, 

\item[] AGS05: meteoritic (for refractories) abundances from \citet{ags05} give 
  $\left(Z/X\right)_{\rm  ph}=0.0165$,

\item[] AGSS09: meteoritic (for refractories) abundances from the most recent 
  determination   of   solar    abundances   by   \citet{agss09}   for   which
  $\left(Z/X\right)_{\rm ph}=0.0178$. One additional model, AGSS09ph, has been
  computed  with the  photospheric  abundances from  \citet{agss09} for  which
  $\left(Z/X\right)_{\rm ph}=0.0181$.
\end{itemize}

\begin{table}[t]
\small
\caption{Adopted solar chemical compositions. \label{tab:compo}}
\begin{tabular}{lcccc}
\hline\hline
& \multicolumn{4}{c}{$\log{\epsilon}$} \\
\cline{2-5}
Elem   &  GS98   &   AGS05\tablenotemark{a}  &
AGSS09\tablenotemark{a} & AGSS09ph\tablenotemark{b} \\
\tableline
C & 8.52 & 8.39 & 8.43 & 8.43 \\
N & 7.92 & 7.78 & 7.83 & 7.83 \\
O & 8.83 & 8.66 & 8.69 & 8.69 \\
Ne & 8.08 & 7.84 & 7.93 & 7.93 \\
Na & 6.32 & 6.27 & 6.27 & 6.24 \\
Mg & 7.58 & 7.53 & 7.53 & 7.60 \\
Al & 6.49 & 6.43 & 6.43 & 6.45 \\   
Si & 7.56 & 7.51 & 7.51 & 7.51 \\
S & 7.20 & 7.16 & 7.15 & 7.12 \\ 
Ar & 6.40 & 6.18 & 6.40 & 6.40 \\ 
Ca & 6.35 & 6.29 & 6.29 & 6.34 \\ 
Cr & 5.69 & 5.63 & 5.64 & 5.64 \\
Mn & 5.53 & 5.47 & 5.48 & 5.43 \\
Fe & 7.50 & 7.45 & 7.45 & 7.50 \\
Ni & 6.25 & 6.19 & 6.20 & 6.22 \\
\hline
\end{tabular}
\tablenotetext{a}{The   adopted   abundances   are   the   recommended   solar
  photo- \newline spheric abundances for  the volatile elements (C, N, O,  Ne
  and \newline Ar) and
  the CI chondritic meteoritic values for the remaining \newline elements. } 
\tablenotetext{b}{The   adopted   abundances   are   the   recommended   solar
  photo- \newline spheric abundances throughout. } 
\tablecomments{Abudances given  as $\log{\epsilon_i}\equiv\log{N_i/N_H}+12$.}
\end{table}


In the new determination of  solar abundances by \citet{agss09}, the difference
between meteoritic  and photospheric abundances is $0.00  \pm 0.04$~dex. While
this agreement is of unprecedented quality, some elements relevant to detailed
solar modelling show larger deviations.  This is particularly the case for Mg,
Ca,  and  Fe for  which  differences  are 0.07,  0.05  and  0.05~dex with  the
photospheric values  being larger in all  cases. To understand  the effects of
these differences in the structure of  the solar interior we have computed two
solar models with  AGSS09 composition, one with meteoritic  and the other with
photospheric   abundances.  In   Table~\ref{tab:compo}  the   relative  number
fractions of  all relevant metals are  given for all sets  of solar abundances
used in the models discussed in this work. 

\section{Results \label{sec:results}}

In what follows, we discuss the most relevant results  related to the interior
structure  of the  Sun for  the four  solar models  described in  the previous
section. Qualitatively,  there is a clear  difference in the  results from the
GS98  model  compared  to  the  others   that  use  a  much  lower  value  for
$\left(Z/X\right)_{\rm ph}$, i.e. AGS05 and AGSS09 models.  This dichotomy has
been widely  discussed in the  literature in connection to models using solar
abundances from \citet{gn93} or \citet{gs98}  on one  hand  and  the
\citet{ags05}  solar  composition on  the other
\citep{bah04,bs05,basu04,turck04,monta04}. Presentation of results is divided
in two sections, the first one comprising general results on solar structure and
inferences from helioseismology, while the second one is devoted to discussion
of solar neutrinos and, briefly, a possible connection with the {\em solar
abundance problem}. 

\subsection{Helioseismology}

\begin{table}[t]
\small
\caption{Main characteristics of solar models.\label{tab:ssm}}
\begin{tabular}{lcccc}
\hline\hline
   &    GS98    &  AGS05    & AGSS09 & AGSS09ph \\
\hline
$Z_{\rm S}$ & 0.0170 & 0.0126 & 0.0134 & 0.0136 \\
$Y_{\rm S}$ & 0.2423 & 0.2292 & 0.2314 & 0.2349 \\
$R_{\rm CZ}/R_\odot$ & 0.713 & 0.728 & 0.724 & 0.722 \\
$\left< \delta c / c\right>$ & 0.0010 & 0.0049 & 0.0038 & 0.0031 \\
$\left< \delta \rho / \rho\right>$ & 0.011 & 0.048 & 0.040 & 0.033 \\
$Y_{\rm c}$ & 0.6330 & 0.6195 & 0.6220 & 0.6263 \\
$Z_{\rm c}$ & 0.0201 & 0.0149 & 0.0160 & 0.0161 \\
$Y_{\rm ini}$ & 0.2721 & 0.2593 & 0.2617 & 0.2653 \\
$Z_{\rm ini}$ & 0.0187 & 0.0139 & 0.0149 & 0.0151 \\
$\alpha_{\rm MLT}$ & 2.15 & 2.10 & 2.09 & 2.12 \\
\hline
\end{tabular}
\end{table}


In Table~\ref{tab:ssm} we summarize the most relevant characteristics of the
solar models used in this work. The first row gives the surface metal mass
fraction $Z_{\rm S}$. The next four rows give quantities that can be directly
tested against helioseismology determinations of solar properties: surface
helium mass fraction $Y_{\rm S}$, depth of the convective envelope $R_{\rm
  CZ}/R_\odot$, and average rms relative differences of the sound speed and
density profiles $\left< \delta c / c\right>$ and $\left< \delta \rho /
\rho\right>$  respectively. $Y_{\rm c}$ and $Z_{\rm c}$ are the present day
central mass fractions of helium and metals and the last three rows give the
initial composition and mixing length parameter, i.e. the free parameters used
to construct an SSM.

Results for the GS98 and AGS05 models are very similar to those already
discussed in the literature (see references above). The updated EOS and some
changes in the nuclear cross sections of a few reactions have very little
impact on the global properties of the models. We describe them here only
briefly. From helioseismology, we know  $Y_{\rm S}=0.2485 \pm 0.0035$
\citep{basu04} and $R_{\rm CZ}/{\rm   R_\odot}=0.713 \pm 0.001$
\citep{basu97}. The GS98 model predicts the right location of the boundary of
the convective envelope and a value of the surface helium mass fraction in
agreement with helioseismology to about 1.8$-\sigma$. On the other hand, the
AGS05 model performance is much worse, giving a 15$-\sigma$ discrepancy for 
$R_{\rm CZ}/{\rm   R_\odot}$ and 5.5-$\sigma$ for $Y_{\rm S}$. Here we note
that only uncertainties from helioseismology are
considered. \citet{montecarlo} have estimated by a series of MonteCarlo
simulations the modelling uncertainties for $R_{\rm CZ}/{\rm   R_\odot}$ 
to be approximately 0.0037 and for $Y_{\rm S}$, coincidentally, 0.0037 as well. 

\begin{figure}[t]
\includegraphics[width=\columnwidth]{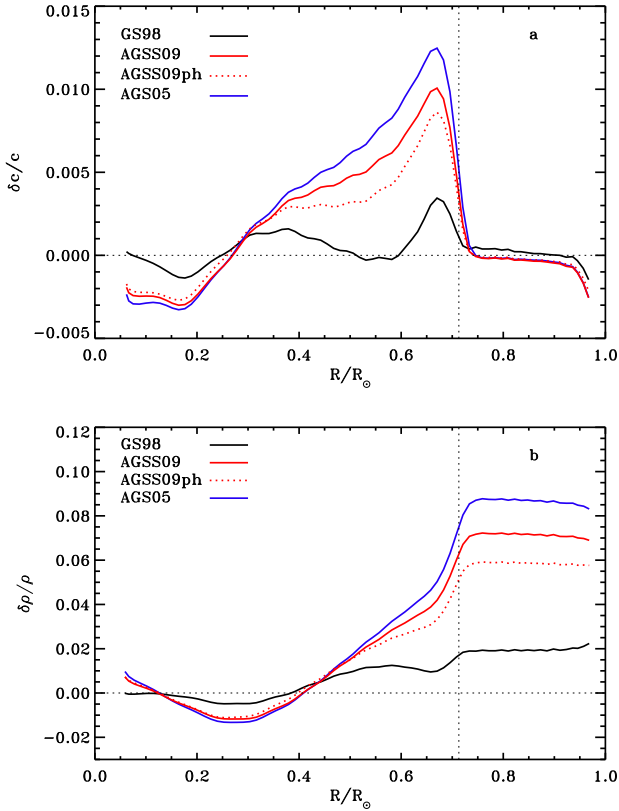}
\caption{Panel {\em  a}: relative sound speed  difference in the  sense (Sun -
  Model)/Model  between  the  solar  sound  speed  profile  as  obtained  from
  helioseismology inversions  and model sound speed profiles.   Panel {\em b}:
  same but for density profiles.   Vertical dotted line in both panels denotes
  the  location of  the bottom  of the  convective envelope  as  inferred from
  helioseismology 
\label{fig:cs}}
\end{figure}

The new solar abundances as determined by \citet{agss09} are slightly higher
than those previously determined by the same group \citep{ags05} (see
Table~\ref{tab:compo}). Changes between 0.03 and 0.05~dex in CNO
abundances are small and, together with the 0.09~dex change in Ne, are not 
able to restore the agreement between solar models and helioseismology.
Although the disagreement is less severe now, 5$-\sigma$ and 11$-\sigma$ for 
$Y_{\rm S}$ and $R_{\rm CZ}/{\rm   R_\odot}$ respectively, it is very large
compared to results of solar models with older \citep{gs98,gn93} solar
abundances. This is also evident when considering the sound speed and density
profiles that are shown in Figure~\ref{fig:cs}. The AGS05 and AGSS09
models give $\left< \delta c / c\right>$ that are 5 and 4 times worse than the
GS98 model respectively. The peak of the discrepancy in the sound speed
profiles is 0.3\% for the GS98 model, 1.2\% for AGS05 and 1\% for AGSS09.  
Analogous results are found for the density profiles. In this case, however,
the larger discrepancies seen in the convective envelopes are associated with
the fact that density inversions include the constraint that the solar mass is
known and, consequently, small differences in density in the innermost region,
at high densities, have to be compensated by larger fractional changes in the
outer much less dense. Both in Table~\ref{tab:ssm} and in Figure~\ref{fig:cs},
it can be seen that the model AGSS09ph, that uses only photospheric abundances
performs better than AGSS09. The reason can be found mostly in the increased
values  of Mg and Fe, by 0.07 and 0.05~dex respectively, in the photospheric
abundances, and not in the overall change in solar metallicity. For example,
Mg contributes to the radiative opacity right the convective envelope and
careful examination of top panel in Figure~\ref{fig:cs} shows that the sound
speed profile of AGSS09ph shows the largest improvement with respect to AGSS09
in the region around 0.6~R$_\odot$. The effect of the increased Fe abundance
can be indirectly appreciated by the initial and surface helium content of the
AGSS09ph model compared to AGSS09. This change results from the relevance of
Fe in the opacity in the central regions and thus in the central most
temperature gradient, which ultimately affects the initial composition of the
model by the condition of fixed solar luminosity imposed on standard solar
models.

\begin{figure}[t]
\includegraphics[width=\columnwidth]{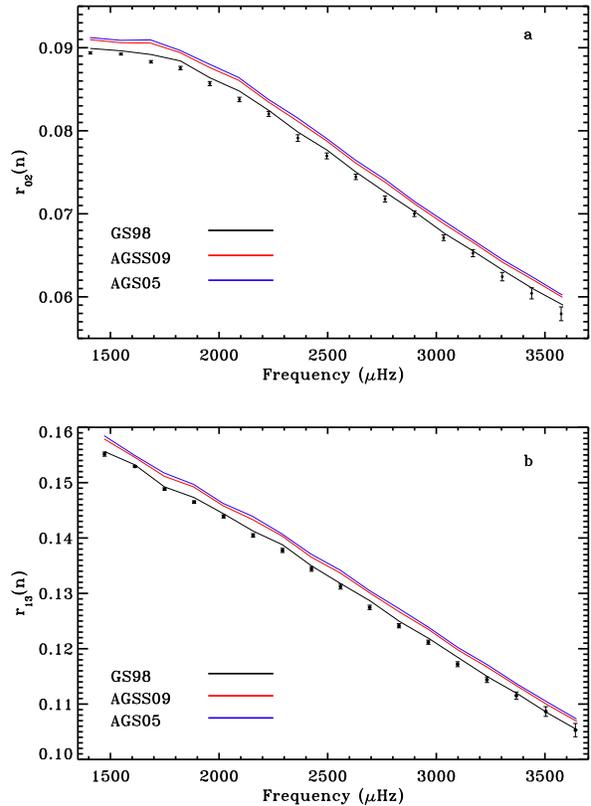}
\caption{Separation  ratios. Comparison between  values determined  from BiSON
  data and  the solar models presented  in this work. Panel  {\em a}: $r_{02}$
  ratios; panel {\em b}: $r_{13}$ ratios
\label{fig:sep}}
\end{figure}

Despite the better performance of the AGSS09ph model, we consider
meteoritic abundances as given in Table~\ref{tab:compo} our preferred
choice. The arguments are that historically, whenever there has been
significant differences between photospheric and meteoritic abundances for
refractory elements, the problem laid in the photospheric
determinations. Also, the excellent overall agreement between the two scales
found by \citet{agss09}, and the lack of accurate atomic data for certain
elements, e.g. for Mg, and the
impossibility of having a homogeneous method of determination of photospheric
abundances for all elements (see discussions in \citealt{agss09} for details)
favor in our view meteoritic abundances as a more secure and robust
option. Hence, our choice.   

Another interesting possibility that global helioseismology offers is to probe
the solar core by means of low-degree ($\ell \leq  3$) modes that penetrate to
the deepest solar regions. In particular, \citet{roxb} have shown that
low-$\ell$ mode frequencies can be used to form the so-called separation
ratios $r_{02}(n)= (\nu_{n,0}-\nu_{n-1,2})/(\nu_{n,1}-\nu_{n-1,1})$ and 
$r_{13}(n)= (\nu_{n,1}-\nu_{n-1,3})/(\nu_{n,0}-\nu_{n-1,0})$, which are
insensitive to the external characteristics of the models. Separation ratios
constructed with very long time series of BiSON data have been used
\citep{bisoni,bisonii} to show that discrepant results between solar structure
inferred from helioseismology and solar models with the AGS05 composition
are not restricted to the outer regions of the Sun (where simplified treatments
of convection by using Mixing Length Theory or similar approaches could in
principle  be  thought of as the culprits of problems in solar modelling) and
extend all the way to the core. In Figure~\ref{fig:sep} we compare the
separation ratios from our models with those ratios derived by \citet{bisoni}
from 4752~days of BiSON data. Results are shown for the GS98, AGS05, and
AGSS09 models and ratios have been connected with lines to help the eye. We
omit results from AGSS09ph model as they practically overlap with those of the
AGSS09 model. As already discussed in \citet{bisoni} and later in
\citet{bisonii}, models with higher metallicity, e.g. from \citet{gs98}, give
good agreement with helioseismology while those using the \citet{ags05}
composition grossly disagree. In this regard, the new AGSS09 model with the
updated solar abundances from \citet{agss09} does not show any improvements
with respect to the AGS05 composition.

Related to  the separation ratios, it is interesting to note a recent work 
\citep{jcd09} where the high sensitivity of the separation ratios and the
small separation 
frequencies (numerators in the definition of separation ratios)
to the properties of the solar core, particularly to the molecular weight
profile, has been used to date the Sun. Interestingly enough, comparison of
the time evolution of the 
separation ratios and small separation frequencies in the S model
\citep{model-S} that uses the \citet{gn93} composition shows the best
agreement with the observed quantities at an age in excellent agreement with
the solar system age as determined from meteoritic samples (see appendix in 
\citealt{bp95}). On the other hand, a solar model with the AGS05 shows the
best agreement with data (which is, nevertheless, much worse than that obtained
with the S model) at an age between 4.8 and 4.9~Gyr.

\subsection{Solar Neutrinos \label{sec:neu}}

For many years, solar neutrinos received a great deal of attention due to the
{\em solar neutrino problem}. With the definite establishment of the
oscillatory nature of neutrinos, the center of attention for studying solar
neutrinos has shifted towards the original goal defined in the early 1960s: to
use neutrinos as a direct probe of how stars shine and of the properties of
the solar core. Currently, two solar neutrino fluxes have been measured
directly: the SNO collaboration has determined directly and with very good
precision the $^8$B flux, an excellent thermometer of the solar core. More
recently, this flux has also been determined by the Borexino
collaboration. More importantly, however, Borexino has been able to determine 
the $^7$Be directly and has already achieved a 10\% accuracy (with 3\% as the
current goal of the collaboration). 

In Table~\ref{tab:neu} the neutrino fluxes predicted by the models used in
this work are given. Results for models with the \citet{gs98} and
\citet{ags05} compositions have already been discussed in the literature,
e.g. \citet{bs05}. We resort here to a qualitative discussion of the
differences. The central temperature in the GS98 model is about 1.2\% higher
than in the AGS05 model and this accounts for the difference ($\sim 20\%$) in
the highly 
sensitive ($\propto (T/T_0)^{16-20}$)~$^8$B flux and also for the reduction in
the $^7$Be flux ($\sim 10\%$) in the AGS05 model. Since solar models 
assume a fixed solar luminosity,
the reduction in nuclear energy released by the ppII chain is compensated by a
slight increase in the rate of ppI chain and the pp and pep fluxes are
slightly increased. 

Larger fractional changes are found for the three fluxes associated with the
CNO bi-cycle. We focus on the $^{13}$N and $^{15}$O fluxes for which
possibilities of direct detection exist in the (relatively) near future with
ongoing efforts by Borexino but mainly with SuperK and SNO+ experiments. These
two fluxes have high sensitive to temperature but are mostly suppressed   
because their rate is directly proportional to the summed abundance of C and N
that is reduced by ($\sim 30\%$) in the \citet{ags05} composition. 

\begin{table}[ht]
\small
\caption{Predicted neutrino fluxes. \label{tab:neu}}
\begin{tabular}{lcccc}
\hline\hline 
Flux\tablenotemark{a}    &   GS98    &  AGS05    & AGSS09 & AGSS09ph \\
\hline
pp & 5.97 & 6.04 & 6.03 & 6.01 \\
pep & 1.41 & 1.44 & 1.44 & 1.43 \\
hep & 7.91 & 8.24 & 8.18 & 8.10 \\
$^7$Be & 5.08 & 4.54 & 4.64 & 4.79 \\
$^8$B & 5.88 & 4.66 & 4.85 & 5.22 \\
$^{13}$N & 2.82 & 1.85 & 2.07 & 2.15 \\
$^{15}$O & 2.09 & 1.29 & 1.47 & 1.55 \\
$^{17}$F & 5.65 & 3.14 & 3.48 & 3.70 \\
\hline 
\end{tabular}
\tablenotetext{a}{Neutrino   fluxes  are   given  in   units   of
  $10^{10}$(pp),   $10^9$($^7$Be), \newline
$10^8$(pep, $^{13}$N, $^{15}$O), $10^6$($^8$B, $^{17}$F) 
and $10^3$(hep)~\hbox{cm$^{-2}$ s$^{-1}$}.} 
\end{table}

With the new \citet{agss09} abundances, neutrino fluxes are very similar to
those from the AGS05 model. In particular, for our preferred meteoritic scale,
the only important differences between these two sets of abundances are the
moderate increase in CNO values that directly affect CNO neutrino fluxes and
the 0.09~dex increase in Ne and the large (0.22~dex) increase in Ar. The last
two elements have some influence on the central temperature of the models and
are responsible for the differences in $^8$B and $^7$Be fluxes between AGS05
and AGSS09 models. Finally, as in the previous section, we also present
results for the 
AGSS09ph model. It is interesting because it illustrates how not only the
overall metallicity of the model is important, but how relative abundances of
elements matter as well. In particular, the larger Fe abundance in the
photospheric abundances account for most of the difference between the fluxes
predicted for this model with respect to the AGSS09 model. 
The interested reader in how individual elements affect neutrino fluxes can
refer to \citet{dnudz} and \citet{bps08}.

\begin{figure}[tb]
\includegraphics[width=\columnwidth]{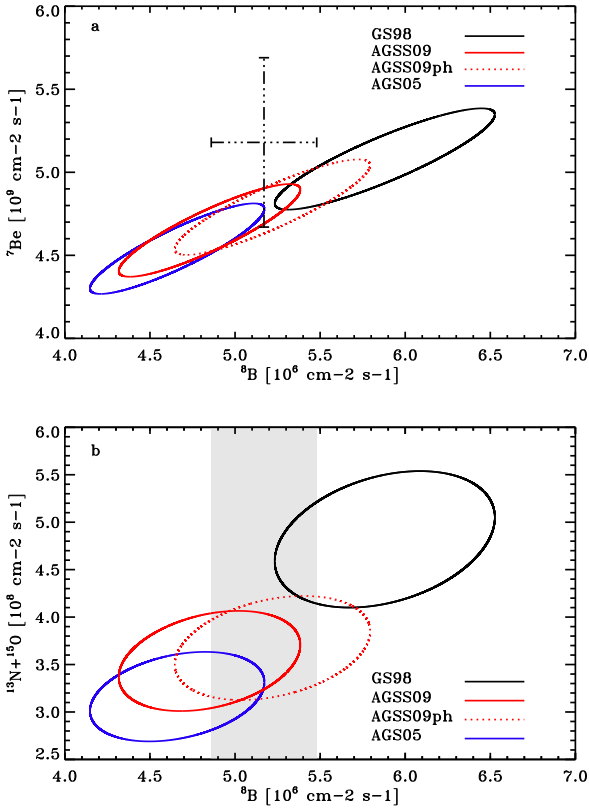}
\caption{Solar  neutrino fluxes.  Panel  {\em a}  shows the  fluxes determined
  directly  from  experiments: $^7$Be  and  $^8$B.  Current measurements  from
  Borexino ($^7$Be)  and the average from  the three phases of  SNO ($8$B) are
  shown  with corresponding  error bars  at $1-\sigma$  level.  Panel  {\em b}
  shows the added contributions of the $^{13}$N and ${15}$O fluxes against the
  $^8$B flux.  The shaded  area represents the  SNO measurement for  the $^8$B
  including $1-\sigma$ uncertainties
\label{fig:neu}}
\end{figure}

The most relevant results for solar neutrinos are summarized in
Figure~\ref{fig:neu}. The top panel shows the $^8$B and $^7$Be fluxes from the
models and the experimental results from SNO ($^8$B) and Borexino ($^7$Be). In
the case of SNO results, because we still lack a joint analysis of the three
different phases, we use the value $\Phi(^8B)= 5.18   \pm  0.29  \times
10^6$~\hbox{cm$^{-2}$ s$^{-1}$} for the flux, a weighed average of the three
phases  \citep{sno1,sno2,sno3} . In the case of Borexino, the measured flux 
after only 192 days of data taking is $\Phi(^7Be)=  5.18 \pm  0.51 \times
10^9$~\hbox{cm$^{-2}$ s$^{-1}$}  \citet{borex}. Model uncertainties for the
fluxes are taken from \citet{bps08}. As discussed in that work, current
neutrino measurements slightly favor results of models with GS98 abundances
over those with AGS05 (or AGSS09). We note, however, that the GS98 and the
AGSS09ph model are at the same level of agreement with helioseismology: GS98
predicts exactly the measured value for $^7$Be and differs from the SNO value
for $^8$B by 1$-\sigma$ (combined model and experimental uncertainties). For
the AGSS09ph model, the situation is exactly opposite. 

In the bottom panel of Figure~\ref{fig:neu}, we show the added $^{13}$N and
$^{15}$O fluxes against the $^8$B flux. Models of different composition
(AGSS09 or GS98) can be more easily distinguished using CNO fluxes as
illustrated in this figure. It should be kept in mind, however, that current
model uncertainties seem to prevent any possibility of a 2$-\sigma$ or better
result. Discussion on main sources of uncertainties and how the situation can
be improved can be found in \citet{bps08}. 

\section{Who is the culprit? \label{sec:opac}}

Since revisions of solar CNO photospheric abundances were strongly revised
downwards \citep{ags05}, attempts have been made to obtain independent
determinations of, or at least to impose constraints on, the solar composition
\citep{basu04,dp06,bisonii}. There has also been a number of works devoted to
analize what changes in the input physics of the models would allow to
construct solar models with low metallicity that are consistent with
helioseismology measurements. We mention, but do not discuss, some of the
proposed changes are: changes in composition, e.g. large enhancement of neon
abundances \citep{ab05,neon,dp06}, enhanced microscopic diffusion
\citep{basu04,monta04,guzik05}, accretion of metal-poor material
\citep{guzik05,castro07}. The viability of these changes and others has been
reviewed at some length by \citet{guzik08} and we refer the interested reader
to that reference for details. The short summary is that none of the proposed
changes can, by itself, offer a solution to the conundrum originated by the
low CNO and Ne abundances presented by Asplund and collaborators. One is left
with the unpleasant option of combining different effects to improve the
helioseismology properties of the low-Z models, and/or to fine tune the
necessary changes in the models. 

Here, we consider in certain amount of detail the effect of new abundances in
the opacities. This has certainly considered before in the light of
\citet{ags05} abundances by a number of authors
\citep{monta04,ab05,bah04,bah05}. More recently, \citet{jcdopac} has
considered this issue by comparing the opacity difference between Model S and
a solar model with solar composition from \citet{ags05}. The authors find
differences in opacities of the order of 30\% at the bottom of the convective
envelope that smoothly decrease towards the center, where 5\% differences are
found. Here, we have followed a very similar line of argument but considered
our GS98 model as the reference model and, in addition to the AGS05 model,
considered those constructed with the two flavors of the new solar abundances,
our AGSS09 and AGSS09ph models. 

\begin{figure}[t]
\includegraphics[width=\columnwidth]{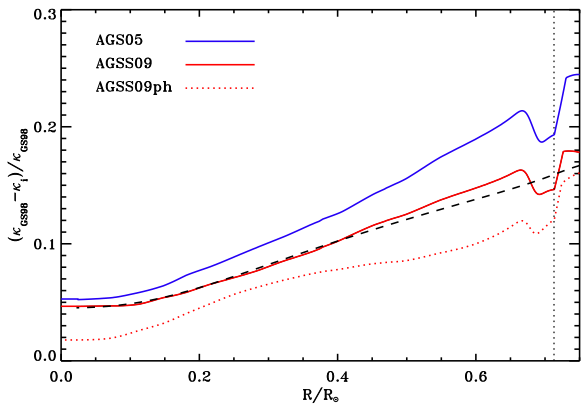}
\caption{Relative opacity deficit of low metallicity solar models relative to
  the GS98 model as a function of radius. Dashed line shows the 
  correction applied  to the  AGSS09 opacities in  the AGSS09+OPAC  model (see
  text).   Vertical dotted  line denotes  the location  of the  bottom  of the
  convective envelope \label{fig:opac}} 
\end{figure}

The main results are shown in Figure~\ref{fig:opac} where we present the
relative differences in opacities between the low-Z models and the GS98
model. For the new AGSS09 model, differences are of the order of 15\%, i.e. a
factor of 2 smaller that those found by \citet{jcdopac}, close to the base of
the convective zone. Somewhat smaller values result in the same region for
the AGSS09ph model due to the larger Mg photospheric abundance. Close to the
center,  the opacity  in the  AGSS09 model  is about  5\% lower  than  in GS98
(similar result as in 
\citealt{jcdopac}) while the AGSS09ph has a deficit of only 2\% due to its
enhanced Fe abundance. We do not open judgement here as whether these
differences are comparable or not to uncertainties in current opacity
calculations. In this regard, we do recall the reader that in the radiative
solar interior, differences between OPAL and Opacity Project opacities do not
rise above 2.5\%, quite below to what is needed to restore the agreement
between solar models with low-Z and helioseismology. 

\begin{figure}[t]
\includegraphics[width=\columnwidth]{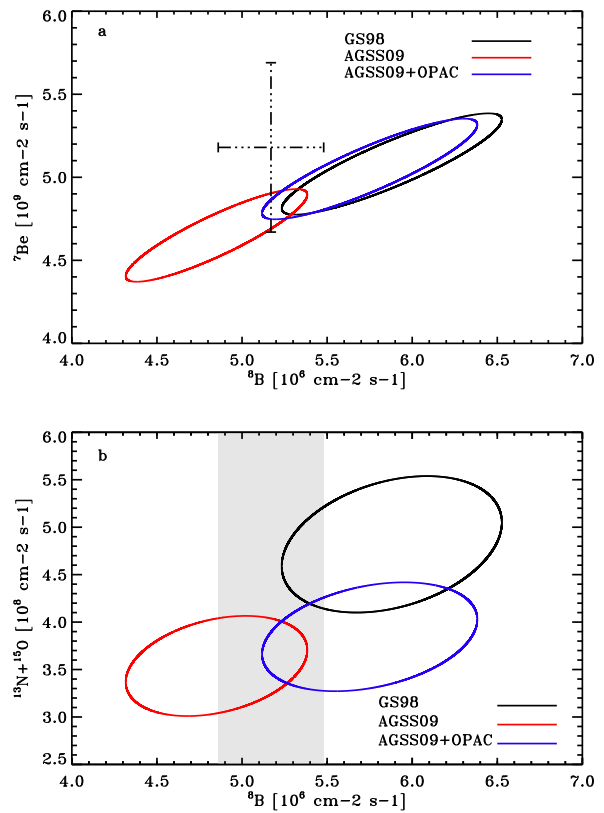}
\caption{Same as Fig.~\ref{fig:neu} for the standard AGSS09 and GS98 models and
a model with AGSS09 composition but  the opacity increased as described in the
text 
\label{fig:neu_opac}}
\end{figure}

In closing this discussion, we present results of a solar model, AGSS09+OPAC,
with the 
same composition as used in the AGSS09 where opacities have been increased by
the fractional amount shown in the black dashed line shown in
Figure~\ref{fig:opac}. \citet{jcdopac} showed that models where the opacity is
increased to compensate changes induced by the modified composition reproduce
all helioseismic properties of the reference model. In this respect, our
results are in the same line; the model with increased opacity performs
equally well than the GS98 model in terms of helioseismic quantities. In
Figure~\ref{fig:neu_opac} we show for GS98, AGSS09, and AGSS09+OPAC models the
results for solar neutrino fluxes. In the top panel it is shown that the
$^7$Be and $^8$B fluxes, affected by the
solar composition mostly through its effect on opacities, of the AGSS09+OPAC
model are very similar to those from the reference GS98 model. There is a
degeneracy in these fluxes between the solar composition and the opacities.
Based on this,
one should be careful when comparing neutrino fluxes from models (with wrong
helioseismic properties, as it is the case for AGSS09) with experimental
determinations: fixing helioseismic properties, e.g. by changing opacities,
will likely affect neutrino predictions.  
On the other hand, in the lower
panel, the added $^{13}$N and $^{15}$O fluxes could in principle be used to
determine the solar core metallicity (more exactly, the total C+N abundance). 
In view of this, we point out the importance of current and future efforts to
measure neutrino fluxes coming from CNO reactions.

\section{Summary \label{sec:conclu}}

We have attempted, in this incomplete review, to describe the current
status of the {\em solar abundance problem} that originated with new
determinations of solar photospheric abundances from Asplund and
collaborators. Results discussed here are based on models computed with both the
original \citep{ags05} and the newest \citep{agss09} solar compositions. Our
reference for a {\em good} solar model is based on the \citet{gs98}
composition. The most important result is that with the new \citep{agss09}
abundances, the qualitative picture that emerged a few years ago, i.e. that
low-Z solar models are in gross disagreement with helioseismology, remains the
same. Quantitatively, the disagreement is less severe because the new
abundances have slightly higher CNO abundances and a somewhat larger Ne
abundance. The changes, however, do not help much neither in restoring the
agreement with helioseismology nor in facilitating the way for alternative
solutions in the form of modified input physics for solar models. We have
described with some detail the effect of the new composition in opacities and
the required change to recover good helioseismic properties. Changes of order
15\% are needed, which are still much higher than currently estimated
uncertainties in radiative opacities for the solar interior. 
In addition to helioseismic properties of the models, we have discussed the
effects of the composition on the predicted neutrino fluxes and compared, when
possible, with results from solar neutrino experiments. Additionally, we have
tried to encourage efforts to experimentally determine neutrino fluxes from
CNO bicycle, since these are the most sensitive fluxes to changes in
abundances of CNO elements, thus offering the best chances for neutrinos to
put direct constraints on the solar core composition.

%
\acknowledgments
I  thank the  people with  whom I  have  been lucky  enough to  work on  solar
modelling  over the  last few  years, particularly  S. Basu,  W.  Chaplin, and
W. Haxton.   My gratitude goes also  to the organizers of  the conference {\em
  Synergies between solar  and stellar modelling}, in particular  Maria Pia di
Mauro,  for  the invitation  to  participate,  the  chosen location,  and  the
exciting atmosphere they contributed to create.

\end{document}